\documentclass[prb,twocolumn,showpacs,superscriptaddress,floatfix]{revtex4}

\usepackage{amsmath}
\usepackage{amsfonts}
\usepackage{amssymb}
\usepackage{bm}
\usepackage{graphicx}

\newcommand{\id}{\mathrm{d}}
\newcommand{\abs}[1]{\left| #1 \right|}
\newcommand{\pder}[2]{\frac{\partial #1}{\partial #2}}
\newcommand{\pderr}[1]{\frac{\partial^2}{\partial^2 #1}}
\newcommand{\psister}{\psi^{\ast}}

\begin{document}

\title{Exciton states in cylindrical nanowires}

\author{A. F. Slachmuylders}
\email{an.slachmuylders@ua.ac.be} \affiliation{Departement Fysica,
Universiteit Antwerpen, Groenenborgerlaan 171, B-2020 Antwerpen,
Belgium}
\author{B. Partoens}
\email{bart.partoens@ua.ac.be} \affiliation{Departement Fysica,
Universiteit Antwerpen, Groenenborgerlaan 171, B-2020 Antwerpen,
Belgium}
\author{W. Magnus}
\email{wim.magnus@imec.be} \affiliation{Interuniversity
Microelectronics Centre, Kapeldreef 75, B-3001 Leuven, Belgium}
\affiliation{Departement Fysica, Universiteit Antwerpen,
Groenenborgerlaan 171, B-2020 Antwerpen, Belgium}
\author{F. M. Peeters}
\email{francois.peeters@ua.ac.be} \affiliation{Departement Fysica,
Universiteit Antwerpen, Groenenborgerlaan 171, B-2020 Antwerpen,
Belgium}


\begin{abstract}
The exciton ground state and excited state energies are calculated
for a model system of an infinitely long cylindrical wire. The
effective Coulomb potential between the electron and the hole is
studied as function of the wire radius. Within the adiabatic
approximation, we obtain `exact' numerical results for the
effective exciton potential and the lowest exciton energy levels
which are fitted to simple analytical expressions. Furthermore, we
investigated the influence of a magnetic field parallel to the
nanowire on the effective potential and the exciton energy.
\end{abstract}

\maketitle

\section{Introduction}
Nowadays, semiconductor nanowires with a small diameter can be
realized in a fast and cost effective manner.  One of the possible
growth techniques, the VLS (vapor-liquid-solid) technique, is a
bottom-up technique which appears to be very
promising\cite{Lieber}. In this method, a liquid metal cluster
acts as a catalyst where reactants dissolve, which leads to the
direct growth from the supersaturated alloy droplet. This growth
method results in a freestanding nanowire that is not surrounded
by any other material, in contrast to the embedded wires.

From a theoretical point of view such semiconductor nanowire
structures are very interesting for understanding the role of
dimensionality on physical properties.  Furthermore, they provide
a great potential for several applications, such as
transistors\cite{trans}, diodes\cite{diode}, memory
elements\cite{memel}, lasers\cite{laser,laser2}, chemical and
biological sensors\cite{sensors}, etc.

These nanowires have a large aspect ratio and can be considered
one-dimensional structures.  They have a length and a diameter in
the micrometer resp.~nanometer range. Electrons and holes are
strongly confined in two directions and there is no confinement in
the direction of the axis of the wire.  Here we will investigate
the exciton binding energy and its lowest excited states. Excitons
form the basis of optical properties reflecting the intrinsic
nature of low-dimensional systems. As a first step we investigate
the potential and energy of the electron-hole pair. Because our
aim is to present numerical `exact' results we will limit
ourselves for the moment, to the \textit{model} system in which
the difference in dielectric constant of the nanowire material and
its surrounding is neglected.  Only in this special case we are
able to present analytical results. Although for `stand alone'
nanowires the dielectric mismatch in the wire (difference in
dielectric constant of the material of the wire and the
environment) is important, the study of it will be postponed for
further investigations.

In previous work\cite{refart}, a similar model system was studied,
but only the ground exciton state was obtained and no systematic
study was presented of the dependence on wire radius, mass of the
carriers, magnetic field and of the excited states. Also impurity
states -- both with and without a magnetic field parallel to the
wire axis -- were considered in previous papers\cite{Chuu,Branis},
with the charged impurity fixed to the cylinder axis and with
averaging performed only over the electronic motion. Furthermore,
the effective potential for the ground state configuration was
calculated before\cite{Banyai} and perfectly fits our results. But
the aim of the present paper is to give a more general result with
different analytical expressions for the effective Coulomb
potential that are fitted to the numerical results.  These fits
haven't been given before and provide us with a tool which is
extremely useful for further numerical calculations. Furthermore,
the binding energy of the exciton is calculated numerically for
arbitrary wire thicknesses, while the results of
Ref.~\onlinecite{refart} were limited to thin wires. Finally, the
addition of a magnetic field is another extension which hasn't
been studied before in detail.  The effective potential was
calculated before for embedded wires\cite{yosyp} for magnetic
fields perpendicular and parallel to the wire axis.  The
techniques used in this paper are the same as in our paper, but in
the paper the focus is not on the effective potential, it is on
the calculation of the exciton energy and comparison with the
experiment.  For high magnetic fields the analytical expressions
as obtained for parabolic confinement\cite{referfc}, can be
accurately fitted to our results. Note that our calculations are
very general, since they apply to arbitrary materials if they can
be described within the framework of the effective mass theory.

This paper is organized as follows.  In
Sect.~\ref{sect:singlepart} we consider the single particle states
of the electron and hole.  In Sect.~\ref{sect:effpot} we calculate
the effective potential of the exciton, using the adiabatic
approximation.  The numerical results for the effective potential
as function of the inter particle distance are fitted to simple
analytical expressions.   The exciton energy is calculated
numerically and semi-analytically in Sect.~\ref{sect:energy}. The
influence of a magnetic field will be investigated in
Sect.~\ref{sect:magnfield} and our conclusions are presented in
Sect.~\ref{sect:conclusion}.

\section{Single particle states}\label{sect:singlepart}
For completeness we first consider the eigenfunctions and
corresponding eigenvalues of a single particle in a
circular\cite{cirkvierk} potential well with radius $R$. Adopting
the effective mass approximation we write the Schr\"odinger
equation as follows:

\begin{equation}
-\frac{\hbar^2}{2m}\vec{\nabla}^2\psi(\vec{r}) +
V(\vec{r})\psi(\vec{r}) = E\psi(\vec{r}),
\end{equation}

where we approximate the confinement potential of the wire by a
hard wall potential

\begin{equation}\label{eq:potential}
V(\vec{r}) = \left \{ \begin{array}{ll} 0 & \mbox{if } \rho \leq R\\
\infty  & \mbox{if } \rho > R \end{array} \right.,
\end{equation}

where $\rho = \sqrt{x^2 + y^2}$. For the freestanding wire the
potential height is given by the work function which is typically
a few eV and therefore very large compared to typical confinement
energies. The wire is oriented along the $z-$axis. Since the
motion in the $z-$direction (free particle in that direction) is
decoupled from the motion in the $(x,y)-$plane and because of
circular symmetry, we have $\psi(\rho,\theta,z) =
F(\rho)e^{-il\theta}e^{ik_zz}$, where $\theta$ is the polar angle.
The Schr\"odinger equation is then reduced to the one dimensional
(1D) equation

\begin{equation}
\rho^2F(\rho)'' + \rho F(\rho)' + ((k^2-k_z^2)\rho^2-l^2)F(\rho) =
0,
\end{equation}

with $k^2 = 2mE/\hbar^2$.  The solution to this equation are the
well known Bessel functions and taking hard wall boundary
conditions into account, one gets for the single particle wave
functions

\begin{eqnarray}\label{eq:besselsol}
\psi_{n,l,k_z}(\rho,\theta,z) & = &
C_{n,l}e^{-il\theta}J_l\left(\frac{\beta_{n,l}}{R}\rho\right)e^{ik_zz},
\end{eqnarray}

and the corresponding eigenenergies

\begin{eqnarray}\label{eq:besselsolen}
E_{n,l,k_z} & = & \frac{\hbar^2 \beta_{n,l}^2}{2mR^2} +
\frac{\hbar^2k_z^2}{2m},
\end{eqnarray}

where $C_{n,l}$ is the normalisation constant and $\beta_{n,l}$ is
the $n^{th}$ order zero of the Bessel function $J_l(x)$.  The zeros
of the Bessel functions are known and can be found in the
literature\cite{abramowitz}: $\beta_{0,0} = 2.4048,\beta_{0,1} =
3.8317,\beta_{0,2} = 5.1356, \beta_{1,0} = 5.5201, \beta_{1,2} =
8.4172, \beta_{2,2}= 11.6198, \ldots$. Notice that the first excited
state is $|0,1,k_z \rangle$ and the second one $|0,2,k_z \rangle$.

\section{Effective exciton potential}\label{sect:effpot}
\subsection{Adiabatic approximation}\label{par:redto1D}
The appropriate Hamiltonian for an exciton in the effective mass
approximation is given by

\begin{eqnarray}
H & = & -\frac{\hbar^2}{2m_e}\vec{\nabla}^2_e + V_e(x_e,y_e)
-\frac{\hbar^2}{2m_h}\vec{\nabla}^2_h + V_h(x_h,y_h) \nonumber \\
&& + V_C\left(\vec{r}_e-\vec{r}_h \right),
\end{eqnarray}

where $m_e$ ($m_h$) is the effective mass of the electron (hole),
$V_e$ ($V_h$) is the confinement potential of the electron (hole)
and

\begin{eqnarray}
&& V_C\left(\vec{r_e}-\vec{r_h} \right) =  \nonumber \\
&& \quad -\frac{1}{4\pi \epsilon}\frac{e^2}{\sqrt{(x_e-x_h)^2 +
(y_e-y_h)^2 + (z_e-z_h)^2}},
\end{eqnarray}

is the Coulomb interaction potential between the electron and the
hole.  In this case, it is convenient to separate the motion of
the two particles into that of the center of mass of the system
and the relative motion, thereby introducing the quantities: $Z
=(m_ez_e+m_hz_h)/(m_e+m_h), z = z_e-z_h, M = m_e+m_h, \mu =
(m_em_h)/(m_e+m_h)$.  Now the Hamiltonian can be rewritten as

\begin{eqnarray}\label{eq:hamil6D}
\lefteqn{H = -\frac{\hbar^2}{2m_e}\vec{\nabla}^2_{x_e,y_e} +
V_e(x_e,y_e) -\frac{\hbar^2}{2m_h}\vec{\nabla}^2_{x_h,y_h} }\nonumber\\
&& + V_h(x_h,y_h)-\frac{\hbar^2}{2M}\pderr{Z}
-\frac{\hbar^2}{2\mu}\pderr{z} \nonumber \\
&& - \frac{1}{4\pi \epsilon}\frac{e^2}{\sqrt{(x_e-x_h)^2 +
(y_e-y_h)^2 + z^2}}.
\end{eqnarray}

In the case of strong lateral confinement, the exciton motion
along the wire is decoupled form the lateral motion of the
particles and we may decouple the wave function as follows

\begin{equation}
\Psi(x_e,y_e,x_h,y_h,z,Z) =
e^{iKZ}\phi(z)\psi(x_e,y_e)\psi(x_h,y_h).
\end{equation}

Because the Coulomb energy in the considered case is much weaker
than the single particle confinement energy, we adopt the
adiabatic approximation, thereby taking $\psi(x,y)$ as the above
obtained single particle states.  Multiplying
Eq.~(\ref{eq:hamil6D}) with $\psister(x_e,y_e)\psister(x_h,y_h)$
from the left and with $\psi(x_e,y_e)\psi(x_h,y_h)$ from the
right, integrating over the lateral coordinates we reduce the
original 6D Schr\"odinger equation for the exciton problem to a 1D
effective Schr\"odinger equation for the relative exciton
coordinate

\begin{equation}
\left( E_e + E_h +
\frac{\hbar^2K^2}{2M}-\frac{\hbar^2}{2\mu}\nabla^2_{z} +
V_{\mbox{eff}}(z)\right)\phi(z) = E_{\mbox{tot}}\phi(z),
\end{equation}

where $E_e$ ($E_h$) are the single-electron (hole) energies, and
the effective exciton potential, $V_{\mbox{eff}}$, is given by

\begin{eqnarray}\label{eq:Veff}
V_{\mbox{eff}}^{n,l}(z) & = & -\frac{e^2}{4\pi \epsilon}\iiiint
\id x_e \id x_h \id y_e \id y_h
\nonumber\\
&& \times
\frac{\abs{\psi_{n,l,k_z}(x_e,y_e)}^2\abs{\psi_{n,l,k_z}(x_h,y_h)}^2}{\sqrt{(x_e-x_h)^2+(y_e-y_h)^2
+ z^2}}.
\end{eqnarray}

Introducing dimensionless units (i.e. length expressed in units of
$R$) and $\alpha = a_B^*/2R$ with $a_B^* = 4\pi \epsilon
\hbar^2/\mu e^2$ the effective Bohr radius, we finally get

\begin{equation} \label{eq:Def1Deqveld}
\left(-\alpha\nabla^2_{z} +
\tilde{V}_{\mbox{eff}}(z)\right)\phi(z) = \tilde{E}_C\phi(z),
\end{equation}

where $z$ is given in units of $R$, $\tilde{V} = V/E_0$,
$\tilde{E_C} = E_C/E_0$, $E_0 = e^2/4 \pi \epsilon R$ and $E_C =
E_{\mbox{tot}}-E_e-E_h-\hbar^2K^2/2M$. To study the bound energy
states of the exciton, we first have to determine the exciton
potential. Examining Eq.~(\ref{eq:Veff}), it is clear that
evaluating this integral numerically is not straight-forward
because of the $1/r$ singularity for $r \rightarrow 0$.

\begin{figure}
\caption{Effective exciton interaction potential as a function of
the electron-hole separation $z$ for three different states of the
electron and hole (Full curve: $n_e = l_e = 0$ and $n_h = l_h = 0$
for the hole; dashed curve: $l_e = l_h = 1$ and $n_e = n_h = 0$;
dotted curve: $l_e = l_h = 2$ and $n_e = n_h = 0$). The regular
Coulomb potential, $1/z$, is also plotted (dashed-dotted) for
comparison. Inset: The effective potential for different sets of
quantum numbers for the electron and hole.} \label{fig:Fig1}
\end{figure}

\begin{figure}
\caption{Density
for the electron and hole in the nanowire for the lowest three
single-particle states.} \label{fig:Fig2}
\end{figure}
The method we used to get rid of this problem, is explained in
Appendix \ref{ap:append1}.  The result of the calculations can be
found in Fig.~\ref{fig:Fig1}, where we show the effective Coulomb
potential when the electron and the hole are in the ground state,
i.e.~$\psi \rightarrow \psi_{0,0}$, ($V^{0,0}$) and in both the
first and second excited state, i.e.~$\psi \rightarrow \psi_{0,1}$
($V^{0,1}$) and $\psi \rightarrow \psi_{0,2}$ ($V^{0,2}$)
respectively (the single particle densities are shown in
Fig.~\ref{fig:Fig2}). Notice how all effective potentials in
Fig.~\ref{fig:Fig1} show the $-1/z$ behaviour for $z \rightarrow
\infty$ and that the effective potential is finite for $z = 0$.

\subsection{Approximate analytical expressions}
Since the calculation of the effective Coulomb potential requires
a considerable amount of computational time, it is highly
desirable to have an analytical expression for it.  Let us first
consider the small $z$ behaviour of the effective potential:

\begin{equation}\label{eq:smallz}
\tilde{V}_{\mbox{eff}}(z) = a + bz + cz^2,
\end{equation}

where $\tilde{V}_{\mbox{eff}}(z)=V_{\mbox{eff}}(z)/E_0$ and $z$ is
in units of $R$. The coefficients are given by

\begin{subequations}
\begin{eqnarray}
a & = & \tilde{V}_{\mbox{eff}}(z = 0) = -2.5961, \\
b & = & \left.\frac{d\tilde{V}_{\mbox{eff}}(z)}{dz}\right|_{z = 0}
=
4.1967, \\
c & = &
\frac{1}{2}\left.\frac{d^2\tilde{V}_{\mbox{eff}}(z)}{dz^2}\right|_{z
= 0} = -5.6140,
\end{eqnarray}
\end{subequations}

for the single particle ground state.  In a similar way, we get $a
= -2.1126$ ($a = -1.9774$), $b = 3.1031$ ($b = 3.0608$) and $c =
-4.5503$ ($c = -5.4277$) for the first (second) excited state.
This polynomial expression, Eq.~($\ref{eq:smallz}$), is
within 1 \% of the numerical result for $z/R < 0.14$.\\
\ \\
Next we will look for an analytical expression that fits the
whole range of $z$-values. Therefore, we have approximated the
full effective potential by Pad\'e approximants. The first Pad\'e
approximation is given by

\begin{eqnarray}\label{eq:Pade1}
\tilde{V}_{\mbox{eff}}(z) = \frac{P_0(z)}{P_1(z)} = \frac{v}{w +
\abs{z}},
\end{eqnarray}

where $P_i(z)$ is a polynomial of $i-$th order.  If we take into
account that $\tilde{V}_{\mbox{eff}}(0) = v/w$ and remember that
for $z \rightarrow \infty$ the effective potential behaves as a
normal Coulomb potential ($-1/z$), we can conclude that $v = -1$
and calculate that $w = 0.3852$ (ground state), $w = 0.4734$
(first excited state) and $w = 0.5057$ (second excited
state).\\
\ \\
This Pad\'e approximation and the result of the calculations are
compared in Fig.~\ref{fig:Fig3}(a) for the ground state and
Fig.~\ref{fig:Fig3}(b) for the first and second excited state.

\begin{figure}
\caption{Plot of
the numerical data and the two Pad\'e approximations,
$P_0(z)/P_1(z)$ and $P_1(z)/P_2(z)$, for (a) the ground state and
(b) the first and second (inset) excited state.} \label{fig:Fig3}
\end{figure}

To improve this result we consider the next Pad\'e approximation:

\begin{eqnarray}\label{eq:Pade2}
\tilde{V}_{\mbox{eff}}(z) = \frac{P_1(z)}{P_2(z)} = \frac{\gamma
\abs{z} + \delta}{z^2 + \eta \abs{z} + \beta}.
\end{eqnarray}

Again, there are certain conditions which impose restrictions to
the values of the parameters.  These conditions are
$\tilde{V}_{\mbox{eff}}(z) \overset{z \rightarrow
\infty}{\longrightarrow}-1/z$, which implies that $\gamma = -1$
and $\delta/ \beta = \tilde{V}_{\mbox{eff}}(z=0)$, which leaves us
with two fitting parameters.  After fitting the data to the
expression in Eq.~(\ref{eq:Pade2}), we find $ \eta = 1.1288 \pm
0.0029, \beta = 0.4705 \pm 0.0015,\delta = -1.2215 \pm 0.0039$ for
the electron and hole in the ground state, $\eta = 1.728 \pm
0.017, \beta = 0.899 \pm 0.011, \delta = 1.900 \pm 0.023$ for the
electron and hole in the first excited state and $\eta = 2.749 \pm
0.061, \beta = 1.559 \pm 0.039, \delta = -1.9774\beta = 3.083 \pm
0.077$ for the electron and hole in the second excited state. The
fitted second Pad\'e approximation is also shown in
Fig.~\ref{fig:Fig3} and we obtain
an excellent fit (i.e.~within 1.5\%) for all states.\\
\ \\
If the confinement potential for the electron and hole is a
parabolic potential, i.e.~$V(x,y) = m\omega_0^2(x^2+y^2)$, it is
possible to perform all integrals in Eq.~(\ref{eq:Veff})
analytically.  The analytical result for the effective potential
was obtained in Ref.~\onlinecite{referfc} and reads

\begin{equation}\label{eq:erfc}
\tilde{V}_{\mbox{eff}}(z) =
-\left(\frac{\pi}{2}\right)^{1/2}\frac{1}{l_0/R}\left[1-\mbox{erf}\left(\frac{\abs{z}}{\sqrt{2}l_0/R}\right)\right]e^{\abs{z}^2/2(l_0/R)^2}
\end{equation}

with the oscillator length $l_0 = (\hbar/m\omega_0)^{1/2}$. It is
remarkable that this potential approaches our effective exciton
potential very closely if we define $l_0$ to be

\begin{equation}\label{eq:erflwde}
l_0/R =
-\frac{1}{\tilde{V}_{\mbox{eff}}(0)}\left(\frac{\pi}{2}\right)^{1/2}.
\end{equation}

i.e.~the expression for the oscillator length extracted from
Eq.~(\ref{eq:erfc}) by putting $z=0$.  It is possible now to
calculate numerically $l_0/R$ for the different effective
potentials: $l_0/R = 0.4828$ (ground state), $l_0/R = 0.5933$
(first excited state) and $l_0/R = 0.6338$ (second excited state).
Using these results in Eq.~(\ref{eq:erfc}), we get an expression
for the effective exciton potential as function of the
interparticle distance. The approximation and the calculated data
can be seen in Fig.~\ref{fig:Fig4} and show a good agreement for
ground state and first excited state (fit within 0.7\% of the
data), while for the second excited state the approximated result
is still reasonable (fits within 3.2\% of the data).

\begin{figure}
\caption{Comparison between the numerical effective electron-hole
potential and the approximate one using the wave function of a
harmonic oscillator for three different sets of quantum numbers of
electron and hole.} \label{fig:Fig4}
\end{figure}

\section{Exciton energy}\label{sect:energy}
\subsection{Numerical calculations}
When a single exciton is created by excitation, its energy is
generally given by $E = E_g + E_e + E_h + E_C$, where $E_g$
denotes the energy band gap, $E_{e(h)}$ is the single electron
(hole) energy and $E_C$ is the (negative) energy of the Coulomb
interaction between electron and hole.  For a given material, we
have to calculate now $E_C$, all other contributions to the
exciton energy are known. Therefore, the 1D Schr\"odinger equation
$\left[\mbox{Eq.}~(\ref{eq:Def1Deqveld})\right]$ was solved using
the finite difference technique. Fig.~\ref{fig:Fig5} shows the
energy of the Coulomb interaction of the three lowest exciton
energy levels as a function of the dimensionless parameter $\alpha
= a_B^*/2R$ for two cases of the effective potential.

\begin{figure}
\caption{Binding
energy as a function of the parameter $\alpha = a^*_B/2R$.  The
solid curves are calculated by using the effective potential
$V^{0,0}$ (electron and hole in the ground state). The symbols
represent the fit which is given by Eq.~(\ref{eq:fitbindener}).
Inset: The same, but now for the effective potential $V^{0,1}$
(electron and hole in the first excited state).} \label{fig:Fig5}
\end{figure}

The binding energy can also be fitted to an analytical expression
now.  Values of $\alpha < 0.1$ are not realistic, since this would
correspond to a large radius and a strong confinement approach
would no longer be valid.  Therefore, the following fit is
suggested for $\alpha$ values exceeding 0.1,

\begin{equation}\label{eq:fitbindener}
\tilde{E}_C = \frac{\xi}{1+b\alpha^\tau}.
\end{equation}

For the ground state configuration of the electron and hole, we
find that $\xi = -2.145 \pm 0.020$, $b = -1.093 \pm 0.022$, $\tau
= 0.5385 \pm 0.0040$, and for the first (second) excited states of
the exciton binding energy we find $\xi = -1.287 \pm 0.031$, $b =
5.42 \pm 0.16$ and $\tau = 0.945 \pm 0.013$ ($\xi = -1.211 \pm
0.016$, $b = 10.23 \pm 0.14$ and $\tau= 0.8830 \pm 0.0034$) (see
Fig.~\ref{fig:Fig5} for the fits). Similarly, for the
configuration where electron and hole are in the first excited
state, we find that $\xi = -1.714 \pm 0.015$, $b = -0.967 \pm
0.020$ and $\tau = 0.5449 \pm 0.0042$ (ground state), $\xi =
-1.0525 \pm 0.028$, $b = 4.56 \pm 0.16$, $\tau = 0.940 \pm 0.016$
(first excited state) and $\xi = -0.965 \pm 0.016$, $b = 8.47 \pm
0.15$, $\tau = 0.8944 \pm 0.0051$ (second excited state).\\
\ \\
From Eq.~(\ref{eq:Def1Deqveld}) it is clear that for small
$\alpha-$values, there is a relatively smaller contribution of the
kinetic energy term $\nabla_z^2 \psi$. As $\alpha$ increases, the
energy of the particle will also increase, because of an increase in
the contribution from the kinetic energy term. For large values of
$\alpha$ the energy levels become closely packed to the
$\tilde{E}_C=0$ level. In fact, because of the $1/z$ behaviour of
the exciton potential for large $z$, the integral $\int{V(z)^{1/2}
\id z} = \infty $ and according to Ref.~\onlinecite{price} there
will be an infinite number of bound states.

\subsection{Comparison with analytical results}
In this section we present an analytical solution of the 1D
Schr\"odinger equation Eq.~(\ref{eq:Def1Deqveld}). Similar
calculations were done earlier by Loudon\cite{loudon}, who solved
the problem of the 1D hydrogen atom (i.e.~the ideal limit of the
1D electron-hole system: infinitesimal wire cross-section and hard
wall confinement) analytically.  To avoid the divergence of the
original potential at $z=0$ Loudon used different models,
e.g.~hard wall conditions at $z=0$, i.e.~$\phi(z=0)=0$. The 1D
Schr\"odinger equation, Eq.~(\ref{eq:Def1Deqveld}), has in general
to be solved numerically. We used the finite difference technique.
In the special case in which the effective potential is
approximated by the first Pad\'e approximant,
Eq.~(\ref{eq:Pade1}), it is possible to obtain an analytical
expression for the wave function. After introducing the new
variable $y = 2\sqrt{-\tilde{E}_C/\alpha}(z+w)$, we can rewrite
Eq.~(\ref{eq:Def1Deqveld}) as the Whittaker equation:

\begin{equation}
\nabla^2_y \phi(y) + \left(-\frac{1}{4} +
\frac{-v}{2y\alpha\sqrt{-\tilde{E}_C/\alpha}} \right) = 0.
\end{equation}

The wave functions are then given in terms of the Whittaker function

\begin{equation}
\phi(y) = Aye^{-y/2} U(1 - \kappa;2;y),
\end{equation}

where $A$ is a normalisation constant and  $\kappa =
-v/2\alpha\sqrt{-\tilde{E}_C/\alpha}$.  The energy is then
obtained by solving the following equations:

\begin{eqnarray}\label{eq:vwdn}
\left. \frac{d \phi(y)}{dy}\right|_{y =
2w\sqrt{-\tilde{E}_C/\alpha}} =0 && \mbox{(even states)}, \nonumber \\
\phi(y = 2w\sqrt{-\tilde{E}_C/\alpha})  =  0 && \mbox{(odd
states)}.
\end{eqnarray}

\begin{figure}
\caption{Ground state (lower curves) and first excited state
(upper curves) energy of the exciton when 1) using the first
Pad\'e approximation for the electron-hole interaction potential
and 2) the numerical expression for the electron and hole in the
ground state.  The curve with symbols is the asymptotic behaviour,
i.e.~Eq.~(\ref{eq:Energnu}).} \label{fig:Fig6}
\end{figure}

We now obtain the energy of the Coulomb interaction
semi-analytically (by solving the transcendent equation) and we
compare it in Fig.~\ref{fig:Fig6} with the numerically calculated
energy. Notice that for large $\alpha$ both energies are close to
each other. But for small $\alpha$-values (which corresponds to
large $R$) there is a underestimation of the exciton energy up to
15 \%. This is not surprising, since the first Pad\'e
approximation leads to a rather poor fit to the effective
potential (see Fig.~\ref{fig:Fig3}(a)). Nevertheless analytical
calculations are useful because they allow for an estimation of
the energy without having to solve the differential equation
(\ref{eq:Def1Deqveld}) numerically.  Furthermore it has to be
noticed that, in the case of large $\alpha$ (which corresponds to
small radii $R$), it is now possible to find a simple analytical
expression\cite{refart} for $\tilde{E}_C$, i.e.

\begin{equation}\label{eq:Energnu}
\tilde{E}_C = -1/4\alpha \nu^2
\end{equation}

where

\begin{eqnarray}\label{eq:nus}
\nu = \nu_m \left \{ \begin{array}{ll} \nu_m  = m + \frac{w}{\alpha} & (\mbox{odd states}) \\
\nu_m = m - \frac{1}{\ln\left(w/m\alpha\right)} & (\mbox{even
states})
\end{array} \right.
\end{eqnarray}

for $m = 1,2,3, \ldots$. These correspond to the bound states of
the exciton.  This asymptotic result is shown in
Fig.~\ref{fig:Fig6} by the square  symbols and is clearly only
valid for large $\alpha$. Note that in the limit of a Coulomb
interaction potential the excited states are twofold degenerate.
The lowest bound state satisfies the special relation

\begin{equation}\label{eq:nu0}
\ln\left( \frac{w}{\nu_0 \alpha}\right) + \frac{1}{2 \nu_0} = 0.
\end{equation}

This approximation is also shown in Fig.~\ref{fig:Fig6}.

\section{Magnetic field dependence}\label{sect:magnfield}
\subsection{Single-particle properties}
When a magnetic field, directed along the wire, is applied, the
single-particle Hamiltonian for an electron ($q = -\abs{e}$) or
hole ($q = \abs{e}$) can be written as

\begin{equation}
H = \frac{1}{2m}  \left(-i\hbar \vec{\nabla} - q\vec{A} \right)^2
+ V(\vec{r}),
\end{equation}

where $V(\vec{r})$ is again the confinement potential of the wire.
Due to the symmetry of the problem we may benefit from the
symmetric gauge $\vec{A} = \frac{1}{2}B\rho\vec{e}_{\theta}$ for
the vector potential. For the same reason, the wave function
$\psi(\rho,\theta)$ is separable, and can be written as
$\psi(\rho,\theta) = R(\rho) e^{-il\theta}$, where the
$z$-dependence has been omitted for the sake of simplicity. The
Schr\"odinger equation now reads

\begin{eqnarray}
&&\left(\pder{^2}{\rho^2}+ \frac{1}{\rho}\pder{}{\rho} -
\frac{l^2}{\rho^2} \right)R +\nonumber \\
&&\left(-\frac{m\omega_c}{\hbar}\mathrm{sgn}(q)l - \left(
\frac{m\omega_c}{2\hbar}\right)^2\rho^2 + \frac{2mE}{\hbar^2}
\right)R = 0,
\end{eqnarray}

where $\omega_c = \abs{e}B/m$ is the cyclotron frequency.
Introducing the following parameters, $ k^{*^2} =
-\mathrm{sgn}(q)l m\omega_c/\hbar+2mE/\hbar^2$ and $1/l_B^2 =
m\omega_c/2\hbar$, and by putting $R(\rho) =
\rho^{\abs{l}}exp(-\rho^2/2l_B^2) f(\rho)$, one gets

\begin{equation}
tf'' + ((\abs{l} + 1) - t)f' - \left(\frac{1}{2}(\abs{l} + 1) -
\frac{l_B^2k^{*^2}}{4}\right)f = 0,
\end{equation}

where the new variable $t$ is given by $t = \rho^2/l_B^2$. The
solution is the confluent hypergeometric function \cite{refgeer}.

\begin{eqnarray}\label{eq:psi}
\psi_{n,l}(\rho,\theta) & = & C_{n,l}
e^{-il\theta}\rho^{\abs{l}}e^{- \frac{
\rho^2}{2l_B^2}} \ _1F_1\left(-a_{n,l};\abs{l} + 1;\frac{\rho^2}{l_B^2}\right), \nonumber \\
\end{eqnarray}

where $C_{n,l}$ is a normalisation constant and

\begin{equation}
-a_{n,l} = \frac{1}{2}(\abs{l} + 1) - \frac{l_B^2k^{*^2}}{4},
\end{equation}

is a constant which is obtained by using the hard wall boundary
conditions, i.e.~$\psi_{n,l}(R,\theta)=0$. The single particle
energy then becomes

\begin{equation}\label{eq:energ}
E_{n,l} = \hbar \omega_c \left(a_{nl} + \frac{1 + \abs{l} +
\mathrm{sgn}(q)l}{2} \right).
\end{equation}

\subsection{Effective exciton potential}

The total Hamiltonian is now given by
\begin{eqnarray}
H_{tot} &=& -\frac{1}{2m_e}\left(\vec{p}_{x_e,y_e} +
\abs{e}\vec{A}\right)^2 + V_e(x_e,y_e) \nonumber \\
&&-\frac{1}{2m_h}\left(\vec{p}_{x_h,y_h} -
\abs{e}\vec{A}\right)^2 + V_h(x_h,y_h) \nonumber \\
&& -\frac{\hbar^2}{2m_e}\nabla^2_{z_e}
-\frac{\hbar^2}{2m_h}\nabla^2_{z_h} \nonumber \\
&& - \frac{1}{4\pi \epsilon}\frac{e^2}{\sqrt{(x_e-x_h)^2 +
(y_e-y_h)^2 + z^2}}
\end{eqnarray}
Again, we can reduce this equation to a 1D effective Schr\"odinger
equation similarly as for the $B=0$ case by: i) separating the
motion of the two particles into the center of mass relative
coordinates, ii) by using the adiabatic approximation for the
exciton wave function, and iii) by introducing the same
dimensionless units as in Sect.~\ref{par:redto1D}. This leads us
to the equivalent of Eq.~(\ref{eq:Def1Deqveld}), where the single
particle energies are now given by Eq.~(\ref{eq:energ}) and the
averaging of the effective potential has to be performed with the
Kummer functions [Eq.~(\ref{eq:psi})] (instead of Bessel
functions). Note that the confluent hypergeometric funtion $\
_1F_1$ is equivalent to the Kummer function $M$:

\begin{equation}
\ _1F_1(a;b;z) = M(a,b,z).
\end{equation}

Fig.~\ref{fig:Fig7} shows the effective potential for different
values of the magnetic field (which corresponds to different
values of the magnetic length $l_B = \sqrt{2\hbar/m\omega_c}$).
\begin{figure}
\caption{Effective exciton potential for different values of the
magnetic field (full lines) in which the electron and hole are in
the ground state. The approximation (open symbols) of
Eq.~(\ref{eq:erfc}) can be used again to fit the calculated
effective potential. For comparison, the effective potential for
the system without magnetic field was also drawn ($l_B/R =
\infty$, dotted line). Note that for this figure, the electron and
hole were in the ground state.} \label{fig:Fig7}
\end{figure}
These potentials can be approximated again by the expression in
Eq.~(\ref{eq:erfc}) (see fit in Fig.~\ref{fig:Fig7}). Therefore it
is more interesting to determine the relation between $l_B$ and
$\tilde{V}_{\mbox{eff}}(z=0)$, since the value of $\tilde{V}
(z=0)$ will enable us to calculate $l_0/R$
(Eq.~(\ref{eq:erflwde})) and therefore to reconstruct the
effective potential for the whole range of $z-$values for an
arbitrary magnetic field. Fig.~\ref{fig:Fig8} shows this curve for
both electron and hole in the ground state and in the first
excited state.
\begin{figure}
\caption{Effective exciton potential at $z = 0$ as function of the
magnetic length $l_B$ (the magnetic field).  The fit is to
Eq.~(\ref{eq:vefffield0}) and
Eq.~(\ref{eq:vefffield1})}\label{fig:Fig8}
\end{figure}\\
\ \\
This figure can be understood as follows. Large values of $l_B$
correspond to the absence of a magnetic field and therefore the
curve converges to the previously calculated value of
$\tilde{V}_{\mbox{eff,}l_B \rightarrow \infty}(z=0)$ for $B=0$.
For small values of $l_B$, Eq.~(\ref{eq:erfc}) matches the
calculated curve, because for high magnetic fields, the
confinement becomes effectively parabolic. Replacing the
oscillator length $l_0$ by $l_B$, the result for large magnetic
fields is recovered:

\begin{equation}
\tilde{V}^{0,0}_{\mbox{eff,}l_B \rightarrow 0}(z=0) =
-\sqrt{\frac{\pi}{2}}\frac{1}{l_B/R}
\end{equation}

Putting together these results into a single formula, we obtain a
fit for the effective potential as function of the magnetic field:

\begin{equation}\label{eq:vefffield0}
\tilde{V}^{0,0}_{\mbox{eff}}(z=0) =
\tilde{V}^{0,0}_{\mbox{eff,}l_B\rightarrow \infty}(z=0)
-\sqrt{\frac{\pi}{2}}\frac{1}{\tilde{l}/R}e^{-pl_B},
\end{equation}

where $1/\tilde{l}^2 = 1/l_0^2 + 1/l_B^2$. Two parameters have
been determined by fitting the data and are $l_0/R = 0.0624 \pm
0.0021$ and $pR = 8.570 \pm 0.090$ for the electron and the hole
in the ground state configuration. Notice that basically we used
the expression for the parabolic confinement Eq.~(\ref{eq:erfc})
and by rescaling $l_0$ to $\tilde{l}$, we include the effect of the magnetic field.\\
\ \\
By making a similar calculation to the one in
Ref.~\onlinecite{referfc}, but for the electron and hole in the
first excited state, we find that

\begin{equation}
\tilde{V}^{0,1}_{\mbox{eff,}l_B \rightarrow 0}(z=0) =
-\frac{11}{16}\sqrt{\frac{\pi}{2}}\frac{1}{l_B/R}
\end{equation}

Therefore we fitted the data for $|n_e,l_e \rangle = |n_h,l_h
\rangle = |n,l \rangle = |0,1 \rangle$ to a slightly different
function,

\begin{equation}\label{eq:vefffield1}
\tilde{V}^{0,1}_{\mbox{eff}}(z=0) =
\tilde{V}^{0,1}_{\mbox{eff,}l_B\rightarrow \infty}(z=0)
-\frac{11}{16}\sqrt{\frac{\pi}{2}}\frac{1}{\tilde{l}/R}e^{-pl_B},
\end{equation}

where $l_0/R = 0.0462 \pm 0.0026$ and $pR = 10.73 \pm 0.17$. The
numerical results together with the results of the fit are shown
in Fig.~\ref{fig:Fig8} and agree very well.

\subsection{Exciton energy}
Previously, we calculated the exciton energy as function of the
dimensionless parameter $\alpha = a_B^*/2R$, which is determined
by the wire radius and the material parameters. In this case,
there is another variable which can be changed and will have an
effect on the energy: the magnetic length $l_B$. Considering the
previous results of the system without magnetic field and knowing
that the shape of the potential does not fundamentally change by
applying a magnetic field (the effective potential well only gets
deeper), we know that also the energy as function of $\alpha$ will
not give a fundamentally different result as before.  Therefore we
calculated the binding energy as function of the magnetic field
(or the magnetic length $l_B$) for a few fixed values of $\alpha$
($\alpha = 10$, $\alpha = 1$ and $\alpha = 0.1$) and the result of
these calculations can be seen in Fig.~\ref{fig:Fig9}.  The main
figure shows only the ground state binding energy, while the inset
shows also the energy of the first and second excited state. The
inset shows that the magnetic field influences the ground state
more than the first and second excited states, which are almost
constant.  For high magnetic fields, which corresponds to $l_B/R
\rightarrow 0$, we find that the effective potential of
Eq.~(\ref{eq:erfc}) becomes a $1/z$-potential\cite{abramowitz}.
This means that we can use Eqs.~(\ref{eq:Energnu})-(\ref{eq:nu0})
again to determine the asymptotic behaviour of $\tilde{E}_C$ as
function of $l_B$.  We find that the ground state energy
$\tilde{E}_C \rightarrow \infty$ for $l_B \rightarrow 0$ for all
values of $\alpha$, whereas for $\alpha=1$ (inset of
Fig.~\ref{fig:Fig9}) the first and second excited state become, in
the limit $B \rightarrow \infty$ degenerate with energy $-1/4$.

\begin{figure}
\caption{Ground
state exciton binding energy for the electron and hole in the
ground state as function of the magnetic length $l_B =
\sqrt{2\hbar/eB}$ for different values of $\alpha$. Inset: the
same but now for the ground state, first and second excited state
with $\alpha = 1$.}\label{fig:Fig9}
\end{figure}

\section{Conclusion}\label{sect:conclusion}
In summary we have calculated the effective interaction potential
and the binding energy for an exciton in a nanowire.  The
effective interaction potential was calculated for different
states of the electron and hole and we were able to present
`exact' numerical results.  In order to reduce the amount of
computational time for future calculations, these results were
used to propose various analytical approximate expressions.  We
fitted the results to three different functions.  A first Pad\'e
approximation gives the correct qualitative behaviour, but the fit
was not optimal. Nevertheless this expression was useful to
perform further analytical calculations to estimate the binding
energy of the exciton, especially for small values of the wire
radius $R$. A second Pad\'e approximation as well as the result
from the parabolic confinement resulted in very good fits.
Furthermore the exciton binding energy was calculated both
semi-analytically and numerically and analytical approximations
were given. We also investigated the influence of a magnetic field
and determined an analytical expression for the effective
potential for arbitrary magnetic field values. Finally we
calculated the binding energy as a function of the magnetic field,
where we found that the ground state binding energy is most
affected by a variation of the magnetic field.

\begin{acknowledgments}
This work was supported by the Flemish Science Foundation
(FWO-Vl), the Belgian Science Policy, the EU network of
excellence: SANDiE and the UA-IMEC, vzw collaborative project.
\end{acknowledgments}

\appendix
\section{Numerical calculation of the effective potential} \label{ap:append1}
To obtain the effective exciton potential of Eq.~(\ref{eq:Veff}),
we have to evaluate four-fold integrals.  They can be rewritten in
polar coordinates as follows, using the same dimensionless units
as before:

\begin{eqnarray} \label{eq:Veffpc}
V_{\mbox{eff}}(z) & =  -R^4\iiiint \rho_e \rho_h\id \rho_e \id
\rho_h \id \theta_e \id \theta_h
& \nonumber\\
& \times
\frac{\abs{\psi_{nl}(R\rho_e,\theta_e)}^2\abs{\psi_{n'l'}(R\rho_h,\theta_h)}^2}{\sqrt{\rho_e^2+\rho_h^2
-2\rho_e\rho_hcos(\theta_e-\theta_h) + z^2}}&
\end{eqnarray}

Due to cylindrical symmetry, we can perform the integral over the
angle $\theta_h$ analytically, yielding

\begin{eqnarray}
\int_0^{2\pi} \id \theta_e \int_0^{2\pi}\frac{ \id
\theta_h}{\sqrt{\rho_e^2+\rho_h^2
-2\rho_e\rho_hcos(\theta_e-\theta_h) + z^2}}
\nonumber \\
= 2\pi \frac{4}{\sqrt{(\rho_e+\rho_h)^2 +
z^2}}\mathbb{K}\left(\frac{4\rho_e\rho_h}{(\rho_e+\rho_h)^2 + z^2}
\right)
\end{eqnarray}

where $\mathbb{K}(m)$ is the elliptic integral of the first kind.
Using a polynomial approximation\cite{abramowitz} for the elliptic
integral,

\begin{eqnarray}
\mathbb{K}(m) & = & \sum_{i = 0}^{4}{a_im_1^i} + \sum_{i =
0}^{4}{b_im_1^i}\ln{\frac{1}{m_1}} \nonumber \\
& \equiv & \mathbb{A}(m_1) + \mathbb{B}(m_1)\ln(1/m_1),
\end{eqnarray}

with $m_1 = 1-m$, we find that

\begin{eqnarray}
&&V_{\mbox{eff}}(z) = -2\pi R^4\int_0^1\id \rho_e \rho_e
\abs{\psi_{nl}(R\rho_e,\theta_e)}^2\nonumber\\
&&\times \underbrace{\int_0^1 4\frac{\id \rho_h \rho_h
\abs{\psi_{n'l'}(R\rho_h,\theta_h)}^2}{\sqrt{(\rho_e+\rho_h)^2 +
z^2}}\left[\mathbb{A}(m_1) + \mathbb{B}(m_1)\ln\frac{1}{m_1}
\right]}_{\equiv I_1} \nonumber
\end{eqnarray}

In a next step, we calculate the integral $I_1$ over $r_h$ using
numerical techniques.  To obtain good convergence, we use the
logarithmically weighted method and consider the following
integral:

\begin{equation}
I(\rho_e,z) = \int_0^1 \id \rho_h F(\rho_h,\rho_e,z)
\ln\left(\frac{(\rho_e-\rho_h)^2 + z^2}{(\rho_e+\rho_h)^2 + z^2}
\right).
\end{equation}

We use the following transformation

\begin{eqnarray} \label{eq:integral}
I(\rho_e,z) & = & \sum_{i = 0}^{s-1} \int_{0}^{h} \id x F(x +
hi,\rho_e,z)
 \nonumber \\
&& \times \ln\left(\frac{(x-(\rho_e-hi))^2 +
z^2}{(x+(\rho_e+hi))^2 + z^2} \right),
\end{eqnarray}

where $h$ denotes the step size and $s$ the total number of steps.
If we replace $F(x+hi)$ by $F_i + (F_{i+1} + F_i)x/h$, we can
write (\ref{eq:integral}) as

\begin{equation}\label{eq:int2}
I(\rho_e,z) = \sum_{i = 0}^{s-1} F_iA_i(\rho_e,z) + \left(F_{i+1}
- F_i\right)C_i(\rho_e,z),
\end{equation}

and the remaining problem is the calculation of the coefficients
$A_i$ and $C_i$. These coefficients are given by

\begin{eqnarray} \label{eq:AC}
A_i(\rho_e,z) & = & \int_{0}^{h} \id x
\ln\left(\frac{(x-(\rho_e-hi))^2 +
z^2}{(x+(\rho_e+hi))^2 + z^2} \right), \nonumber \\
C_i(\rho_e,z) & = & \frac{1}{h}\int_{0}^{h} \id x x
\ln\left(\frac{(x-(\rho_e-hi))^2 + z^2}{(x+(\rho_e+hi))^2 + z^2}
\right).
\nonumber \\
\end{eqnarray}

These integrals can be performed analytically which leads to the
following results:

\begin{subequations}
\begin{eqnarray}
A_i(\rho_e,z) & = & a(h-(\rho_e-hi)) - a(h+(\rho_e+h_i)), \\
C_i(\rho_e,z) & = & h^{-1}\left[c(h-(\rho_e-hi)) -
c(h+(\rho_e+h_i))\right]
\nonumber \\
&& - 2r_e,
\end{eqnarray}
\end{subequations}

with

\begin{widetext}
\begin{eqnarray}
a(y) & = & H_i(y) + H_i(h-y) +
2z\left[\arctan\left(\frac{y}{z}\right) +
\arctan\left(\frac{h-y}{z}\right) \right], \nonumber \\
H_i(y) & = & y\ln(y^2+z^2), \nonumber \\
c(y) & = & \frac{1}{2} \left(y(2h-y) + z^2 \right)G_i(y) +
\frac{1}{2} \left((h-y)^2 - z^2 \right)G_i(h-y)
2z(h-y)\left[\arctan\left(\frac{y}{z}\right) +
\arctan\left(\frac{h-y}{z}\right) \right], \nonumber \\
G_i(y) & = & \ln(y^2+z^2).
\end{eqnarray}
\end{widetext}

Now we can write, using the logarithmically weighted method,

\begin{eqnarray} \label{eq:intrh}
I_1 & = & \int \frac{4\id \rho_h \rho_h
\abs{\psi_{n'l'}(R\rho_h,\theta_h)}^2}{\sqrt{(\rho_e + \rho_h)^2
+ z^2}}\left(\mathbb{A}(m_1) + \mathbb{B}(m_1)\ln(m_1) \right) \nonumber \\
& = & \sum_{i=0}^{s-1} \int_0^h \id x F(x+hi)\left(\mathbb{A}(m_1) + \mathbb{B}(m_1)\ln(m_1) \right) \nonumber \\
& = & \sum_{i=1}^{s-1}\left(F_ih \mathbb{A}(m_1) + F_i \left(A_i +
C_{i-1} - C_i \right)\mathbb{B}(m_1)\right) \nonumber \\
&& + F_s\frac{h}{2} \mathbb{A}(m_1) + F_sC_{s-1} \mathbb{B}(m_1),
\end{eqnarray}

where

\begin{equation}
F_i \equiv F(hi) = \frac{4hi
\abs{\psi_{n',l'}(Rhi,\theta_h)}^2}{\sqrt{(\rho_e + hi)^2 + z^2}}.
\end{equation}

The final step is to calculate then the integral over $\rho_e$.
Similarly, we can also perform this last integration over
$\rho_e$. Using the above mentioned result of
Eq.~(\ref{eq:intrh}), we obtain the final expression for the
effective potential in the $z$-direction:

\begin{widetext}
\begin{eqnarray}
V_{\mbox{eff}}(z) & = & -\frac{2\pi}{N^2}R^4 \Bigg[h \sum_{j =
1}^{s-1} hj\abs{\psi_{nl}(Rhj,\theta_e)}^2 \Big[\big[S_1(hj)h +
T(hj)\frac{h}{2}\mathbb{A}(m_1)\big] - \big[S_2(hj) +
T(hj)C_{s-1}\mathbb{B}(m_1)\big] \Big]  \nonumber \\
&& + \frac{h}{2}hs\abs{\psi_{nl}(Rsh,\theta_e)}^2
\Big[\big[S_1(hs)h + T(hs)\frac{h}{2}\mathbb{A}(m_1)\big] -
\big[S_2(hs) + T(hs)C_{s-1}\mathbb{B}(m_1)\big] \Big] \Bigg]
\nonumber
\end{eqnarray}
\end{widetext}

with

\begin{eqnarray}
S_1(x) & = &
\sum_{i=1}^{s-1}\frac{4hi\abs{\psi_{n'l'}(Rhi,\theta_h)}^2}{\sqrt{(x
+ hi)^2 + z^2}}\mathbb{A}(m_1), \\
T(x) & = & \frac{4sh\abs{\psi_{n'l'}(Rsh,\theta_h)}^2}{\sqrt{(x +
sh)^2 + z^2}}, \\
S_2(x) & = &
\sum_{i=1}^{s-1}\frac{4hi\abs{\psi_{n'l'}(Rhi,\theta_h)}^2}{\sqrt{(x
+ hi)^2 + z^2}}(A_i + C_{i-1}-C_i)\mathbb{B}(m_1). \nonumber \\
\end{eqnarray}


\begin{thebibliography}{99}
\bibitem{Banyai}L. Banyai, I. Galbraith, C. Ell, and H. Haug, Phys.
Rev. B \textbf{36}, 6099 (1987).  Note that this paper contains
several mistakes in the formulas; the figures, however, are
correct which indicates that mistakes are typing errors.
Nevertheless, one should be carefull and not just copy the
formulas.
\end{thebibliography}
\end{document}